\documentclass[aps,prb,reprint,superscriptaddress]{revtex4-2}
\usepackage{graphicx}	
\usepackage{siunitx}

\usepackage{xcolor}
\usepackage{amsmath}
\usepackage{amssymb}
\usepackage{hyperref}
\hypersetup{
	colorlinks=false,
	linkcolor=blue,
	urlcolor=blue,
	citecolor=blue,
	pdfpagemode=FullScreen,
}
\usepackage{printlen}
\usepackage[capitalise]{cleveref}

\usepackage{enumitem}
\usepackage{float}
\usepackage{lipsum}
\makeatletter
\def\maketitle{
	\@author@finish
	\title@column\titleblock@produce
	\suppressfloats[t]}
\makeatother
\usepackage{setspace}

\begin{document}
\title{Bulk Thermal Conductance of the $5/2$ and $7/3$ Fractional Quantum Hall States in the Corbino Geometry}
\author{F. Boivin}
\affiliation{Department of Physics, McGill University, Montréal, Québec, H3A 2T8, Canada}
\author{M. Petrescu}
\affiliation{Department of Physics, McGill University, Montréal, Québec, H3A 2T8, Canada}
\author{Z. Berkson-Korenberg}
\affiliation{Department of Physics, McGill University, Montréal, Québec, H3A 2T8, Canada}
\author{K. W. West}
\affiliation{Department of Electrical Engineering, Princeton University, Princeton NJ 08544 USA}
\author{L. N. Pfeiffer}
\affiliation{Department of Electrical Engineering, Princeton University, Princeton NJ 08544 USA}
\author{G. Gervais}
\email{gervais@physics.mcgill.ca}
\affiliation{Department of Physics, McGill University, Montréal, Québec, H3A 2T8, Canada}

\begin{abstract}
		In this work, making use of time-resolved \textit{in situ} Joule heating of a two-dimensional electron gas (2DEG) in the Corbino geometry, we report bulk thermal conductance measurements for the $\nu = 5/2$ and $\nu = 7/3$ fractional quantum Hall (FQH) states for electron temperatures ranging from 20 to 150 mK. We compare our findings with a recent study by Melcer \textit{et al.} [\href{https://www.nature.com/articles/s41586-023-06858-z}{Nature \textbf{625}, 489 (2024)}] that observed a finite bulk thermal conductivity $\kappa_{xx}$ in FQH states. In spite of the large size difference and the vastly different experimental schemes used to extract $\kappa_{xx}$, we find in large part that both experiments yield similar results and conclude that the bulk of FQH states thermally conducts and violate the Wiedemann-Franz law by a wide margin. Slight discrepancies between both studies are further discussed in terms of particle-hole symmetry in the vicinity of the 5/2 and 7/3 FQH states.
\end{abstract}

\date{\today}
\pagenumbering{arabic}

\maketitle

\setlength{\parskip}{1em}

\par \textit{Introduction.} The first observation of the fractional quantum Hall effect \cite{Tsui1982} led to a revolution in our understanding of strongly correlated electrons in 2DEG systems. Since then, this platform has led to numerous theoretical and experimental investigations of two-dimensional quantum states of matter such as composite fermions (CF) \cite{Jain1989,Halperin1993}, and subsequently, to possible non-Abelian anyons \cite{Moore1991} following the observation of an even FQH state (\textit{i.e.}, $\nu=5/2$ \cite{Willett1987}) in the second Landau level. The latter is particularly puzzling and it has been the subject of extensive experimental studies \cite{Eisenstein1988, Dolev2008, Samkharadze2011, Tiemann2012} motivated in part by its potential for fault-tolerant quantum computations with non-Abelian anyonic quasiparticles \cite{Nayak2008}. 

Nonetheless, stepping away from transport and interferometric measurements, thermal transport is a promising avenue since it provides access to new thermodynamic information that could test predictions for non-Abelian topological order of certain FQH states. Notably, in the clean limit, the longitudinal thermopower ($S_{xx}$) can be linked to the entropy  \cite{Cooper2009,Yang2009,Chickering2013,Liu2018} and potentially could provide insights on the non-Abelian nature of FQH states. In addition, the measurements of the thermal Hall conductance can potentially shed light on the topological order and propagating edge modes in FQH states \cite{Kane1997}. This has been studied for several FQH states in both GaAs/AlGaAs \cite{Banerjee2017} and graphene systems \cite{Srivastav2022} where quantized values of thermal Hall conductivity $\kappa_{xy}$ were observed \cite{Banerjee2018,Dutta2021,Dutta2022}, \textit{i.e.} $\kappa_{xy}T = \nu_{Q} \kappa_{0} T$, where $\nu_{Q}$ depends on the topological order and $\kappa_{0}=\frac{\pi^2 k_{B}^2}{3h}$ is the thermal conductance quantum. 

The situation contrasts greatly in the case of the longitudinal thermal conductivity $\kappa_{xx}T$ in the bulk, as only a few experimental studies \cite{Venkatachalam2012,Melcer2024} have been able to measure the bulk contribution of $\kappa_{xx}$. Attempting to isolate the \textit{purely bulk} contribution from edge modes represents a great challenge. As demonstrated experimentally recently by Melcer \textit{et al.} \cite{Melcer2024}, the bulk is not straightforwardly related by a Wiedemann-Franz law and as a result is not necessarily thermally insulating. In fact, in their studies, they reported a finite bulk thermal conductance in the 5/2 and 7/3 FQH states and rationalized this finding \textit{via} a theoretical model for the localized states present in the bulk. 

Here, the longitudinal thermal conductance $K$ was obtained in the Corbino geometry, which has no edge states and thus is inherently bulk. Successful past experiments in the Corbino configuration demonstrated its suitability to probe thermodynamic properties in the 5/2 and 7/3 FQH states \cite{Schmidt2015,Schmidt2017,Petrescu2023}. In this work, we observe that even though the Wiedemann-Franz law is not respected in absolute value, in accordance with the work of Melcer \textit{et al.}, the trend of $K$ with respect to the magnetic field is in agreement with the electrical conductance and differ in trend with Ref. \cite{Melcer2024}.

\begin{figure}[!thb]
	\centering
	\includegraphics[width=0.45\textwidth]{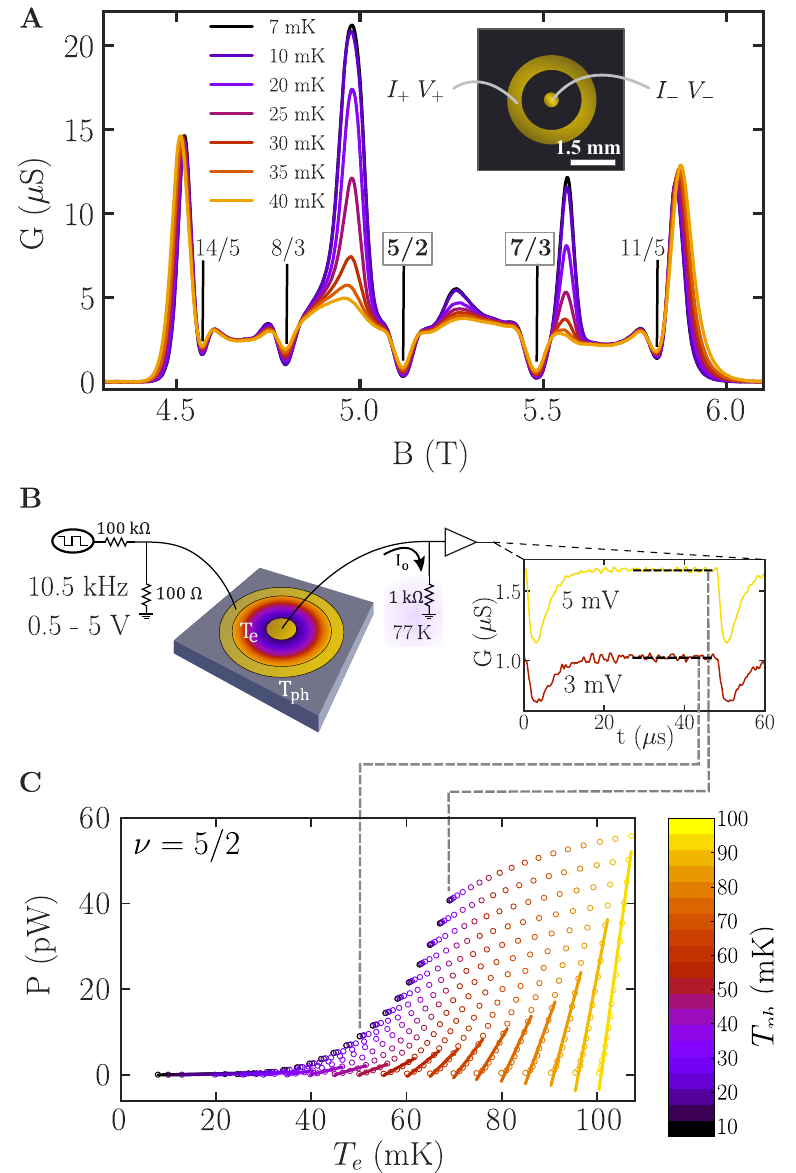}
	\caption{\textbf{(A)} Corbino conductance as a function of magnetic field in the second Landau level, where the most well-developed FQH states are indicated by arrows. A schematic of the high mobility 2DEG Corbino device (sample CB05) is shown in the inset (to scale). \textbf{(B)} Circuit diagram and schematics of the experimental setup and data analysis procedure used to extract the thermal conductance. A unipolar square wave is applied to the Corbino device, leading to Joule heating of the 2DEG. The resulting change in conductance after equilibrium is recorded and used as an \textit{in situ} thermometer to relate the increase in electron temperature $T_e$ with respect to the input power $P$. \textbf{(C)} The data shows the power required to heat the 2DEG to an electron temperature $T_e$ at a given phonon bath temperature $T_{ph}$. The slopes of $P$ versus $T_e$, shown in a color with respect to phonon temperature, are proportional to the thermal conductance up to a geometric factor (see main text of Ref. \cite{Schmidt2019}) in the low-power limit. }\label{fig:1} 
\end{figure}

\begin{figure*}[!thb]
	\centering
	\includegraphics[width=0.9\textwidth]{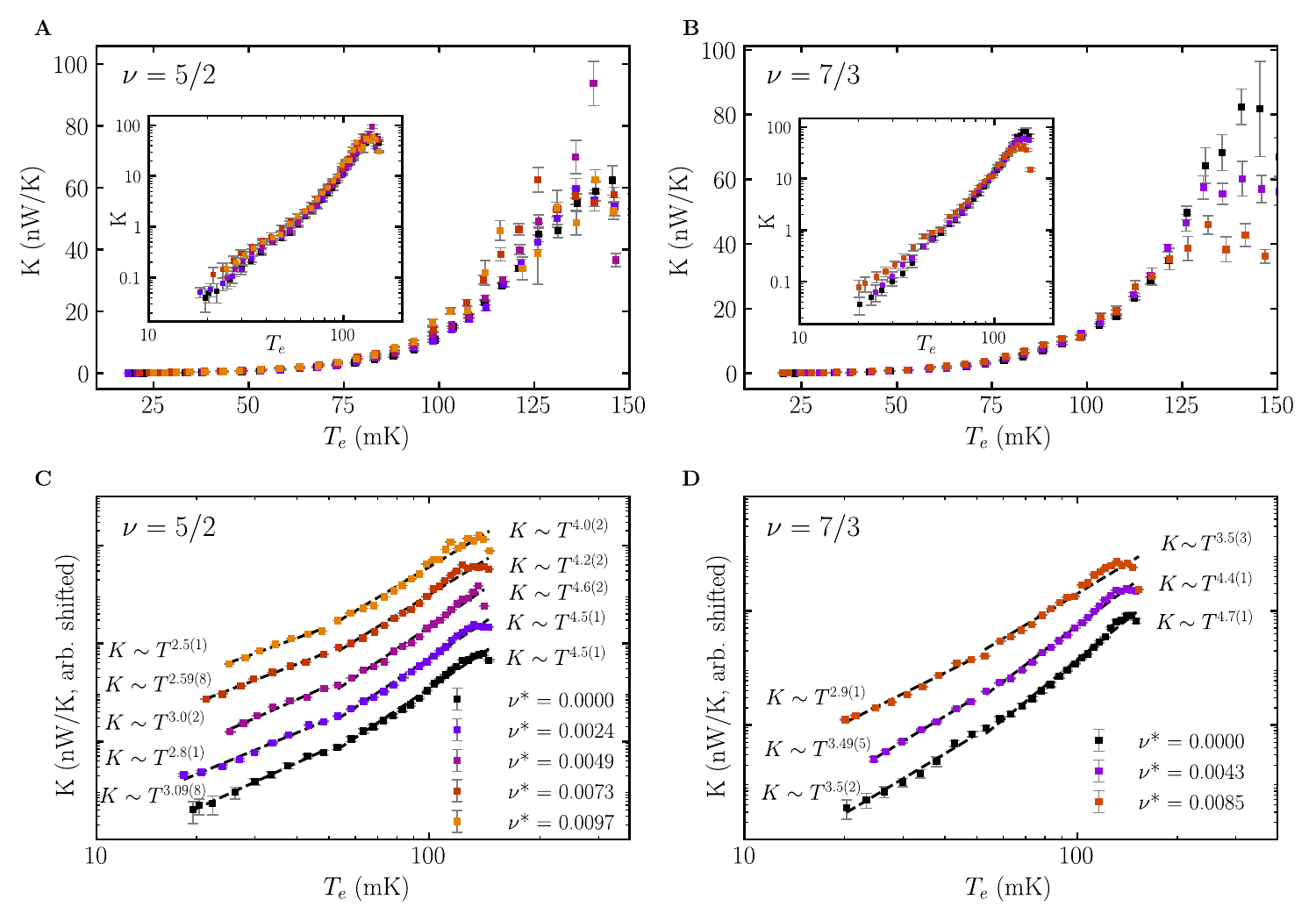} %
	\caption{Temperature dependence of the thermal conductance in \textbf{(A)} for the $\nu=5/2$ FQH state and \textbf{(B)} the $\nu=7/3$ state. The same data is shown in the inset on a log scale. \textbf{(C), (D)} Thermal conductance data is shown on a log scale with each curve arbitrarily shifted vertically to highlight the individual behaviors of $K(T_e)$ at filling factor $\nu^*$. A power law fit is shown as dashed lines, below and above 50 mK. }\label{fig:2}
\end{figure*}

\par \textit{Methods.} The device measured is shown in the inset of Fig. \ref{fig:1}\textbf{A}. It is patterned on a GaAs/AlGaAs heterostructure having a quantum well width of 30~nm, an electron density of $n_e = 3.08(1) \times \mathrm{10^{11}~cm^{-2}}$, and a measured mobility of $22(2) \times 10^{6}~\mathrm{cm^2/V \cdot s}$  \cite{Schmidt2015,Schmidt2017,Schmidt2019}. The inner circular contact has a radius of 0.25 mm, whereas the outer ring contact has an inner and outer radius of 1 mm and 1.5 mm, respectively. The contacts consist of Ge/Ni/Au layers patterned by photolithography. Further details pertaining to the fabrication and characterization of the device can be found in Refs. \cite{Schmidt2015,Schmidt2019}. Finally, during the initial cooling process, the 2DEG mobility was optimized by illumination with a red LED down to a temperature of $\sim$ 5 K.

The experiment was performed in a BlueFors LD-250 dilution refrigerator with a base temperature of 7 mK. The Corbino 2DEG was thermalized using thermocoax cables together with a sequence of radiation shields that allowed us to reach an electron temperature below 20 mK. Two conduction measurement circuits were employed to acquire the electrical and thermal transport properties of the device in a magnetic field. For electrical conductance, the circuit is schematically shown in the inset of Fig. \ref{fig:1}\textbf{A} and is detailed further in the supplementary material. In this circuit configuration, an AC voltage bias of 0.1 mV at 13.333$~$Hz was applied to the device contacts using an SR830 lock-in amplifier and a voltage divider circuit. The resulting current flowing through the device was detected through the induced potential difference in a 1$~$k$\Omega$ resistor in series with the same SR830 lock-in amplifier.

\begin{figure*}[!thb]
	\centering
	\includegraphics[width=\textwidth]{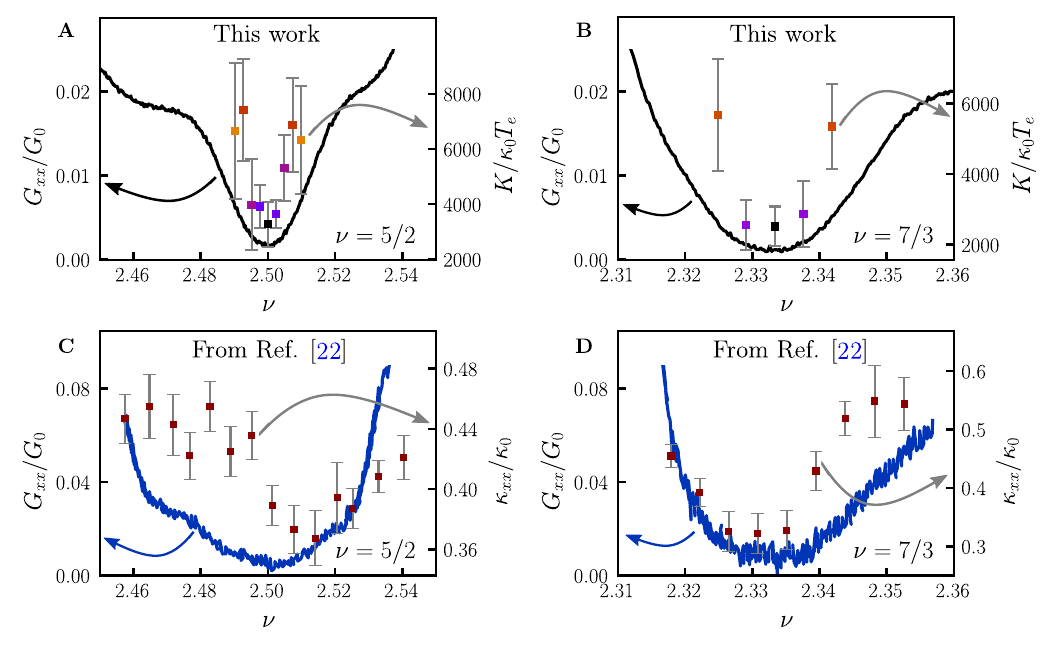} %
	\caption{\textbf{(A), (B)} The electrical conductivity $G_{xx}/G_0$ (left axis) and thermal conductivity $K/\kappa_0 T_e$ (right axis) shown for \textbf{(A)} $\nu=5/2$ and  \textbf{(B)} $\nu=7/3$ FHQ states as a function of filling factor. The color coding on data points corresponds to the same partial filling factors in absolute value ($|\nu^*|$) that are presented in Fig. \ref{fig:2}\textbf{C,D}. \textbf{(C), (D)} The same quantities ($G_{xx}$,$\kappa_{xx}$) as in \textbf{(A), (B)} digitized from the work of Melcer \textit{et al.} \cite{Melcer2024} for the same FQH states as in \textbf{(A), (B)}. Note that Ref. \cite{Melcer2024} probed the thermal conductivity $\kappa_{xx}/\kappa_{0}$  directly whereas we measure the thermal conductance $K$ which was then converted to the thermal conductivity $K/\kappa_{0}T$. }\label{fig:3}
\end{figure*}

The experimental procedure used to determine thermal conductance is illustrated in Fig. \ref{fig:1}\textbf{B,C}. As detailed in Ref. \cite{Petrescu2023}, the temperature gradient was generated by applying a 10.5 KHz unipolar square wave excitation ranging from 0.5 to 5 mV between the Corbino device contacts. This results in Joule heating of the 2DEG, and it induces a temperature gradient between the inner and outer contacts. The resulting change in electrical conductance (see Fig. \ref{fig:1}\textbf{B}) is measured by a Zurich HF2LI lock-in amplifier that reads the voltage across a 1 k$\Omega$ resistor submerged in liquid nitrogen for noise reduction. Upon correcting for wiring resonance transients using the \textit{shift and subtract} method \cite{Schmidt2019}, the equilibrium value of the conductance $G_{eq}(T_e)$ is then used to infer the electron temperature $T_e$ reached at a given power $P$ and phonon bath temperature $T_{ph}$, as shown by the dashed lines connecting $G_{eq}$ in Fig. \ref{fig:1}\textbf{B} to the corresponding data points in Fig. \ref{fig:1}\textbf{C}. A linear fit was then performed in the adequate regions of the $P(T_e)$ curves for each phonon temperature $T_{ph}$ to extract the thermal conductance $K(T_e)$. In the Corbino geometry the radial temperature gradient is non-linear and as a result the thermal conductance $K$ needs to be corrected by a geometric factor $K = 1.83K_{app}$, where $K_{app}$ is the apparent thermal conductance assuming a linear gradient (see Ref. \cite{Schmidt2019} for more details). Finally, we note that thermal conductance $K$ for the analysis related to Fig. \ref{fig:2}, will be converted to thermal conductivity in Fig. \ref{fig:3} in order to compare our results with those of Ref. \cite{Melcer2024}.

\par \textit{Results.} The main findings of this work are shown in Fig. \ref{fig:2}, where Fig. \ref{fig:2}\textbf{A} and Fig. \ref{fig:2}\textbf{B} show the thermal conductance as a function of temperature in a linear and log scale (inset) for the 5/2 and 7/3 FQH states, respectively, and for a selected range of non-exact filling factors $\nu^*$. Here, $\nu^*$ is defined as a small deviation from the exact filling factor of a given FQH state $\nu_0$, \textit{i.e.} $\nu^*\equiv [\nu-\nu_0]$, where $\nu$ is the filling factor at a given magnetic field. Note that above 125 mK, the uncertainty and the exactitude of the data points deteriorate, mainly because this temperature range becomes close to the energy gap of the 5/2 and 7/3 FQH states, which were determined to be 172(5) mK and 207(5) mK, respectively \cite{Petrescu2023}. A closer inspection of the thermal conductance displayed on a log scale in the insets of Fig. \ref{fig:2}\textbf{A} and Fig. \ref{fig:2}\textbf{B} reveals the presence of a weak crossover around 50 mK. This crossover behavior was investigated in more detail and is shown in Fig. \ref{fig:2}\textbf{C} and Fig. \ref{fig:2}\textbf{D}, with power law curve fits above and below 50 mK. Note that the thermal conductance data points were arbitrarily vertically shifted for clarity and the unshifted data are shown in the inset of Fig. \ref{fig:2}\textbf{A,B}. From this analysis, we observe a change in the power scaling roughly from $T^4$ to $T^{3}$ below 50 mK for all partial filling factors investigated in this work. This is consistent with previous experiments in the same device \cite{Schmidt2017,Schmidt2019} where $K$ was found to follow a $T^4$ power law, which can be attributed to phonon emission cooling \cite{Schmidt2019,price1982,karpus1988,Appleyard1998,Mittal1996}. Nevertheless, we are unable to pinpoint the exact mechanism at the origin of the crossover observed in the bulk thermal conductance of the 5/2 and 7/3 FQH states that is observed below 50 mK in our work. 

\textit{Discussion.} Owing to the interest in the thermal properties of the 5/2 FQH state for its possible non-Abelian nature that could give rise to edge currents quantized as half-integer \cite{Banerjee2018}, we opted to compare our findings with a recent study from Melcer and co-workers \cite{Melcer2024}. In their work, they performed analogous measurements based on a Hall bar device with grounded edge modes to extract the bulk thermal conductivity of the 5/2 and 7/3 FQH states. Importantly, we point out that there are crucial distinctions in the methodology of Melcer \textit{et al.}'s \cite{Melcer2024} and our work. In the former, the electronic and heat transport occurs over a few micrometers, and the bulk contribution is obtained \textit{via} grounding the edge channels while measuring locally the propagating power. Due to the noise-based measurement of the impinging heating power, Melcer \textit{et al.}'s \cite{Melcer2024} measurements are more naturally presented in terms of the thermal conductivity $\kappa_{xx}$ (\textit{i.e.}, units of W/K$^{2}$) and consequently, $\kappa_{xx}$ is further normalized by the quantum of thermal conductance $\kappa_0 = 9.464\times 10^{-13} $  W/K$^{2}$ to report a unit-less quantity that provides information on the magnitude of the bulk thermal conductance. In our case, to properly compare our data with their results, we converted our raw thermal conductance $K$ to the unit-less thermal conductivity by dividing with the electron temperature, the \textit{number of squares} in our geometry $\frac{\log(r_2/r_1)}{2\pi}$, and then by $\kappa_0$, as shown in Fig. \ref{fig:3}\textbf{A,B}. The conductance data from Ref. \cite{Melcer2024} was digitized from the article figures. Note that the electrical conductance measured in the Corbino geometry was converted to conductivity using the \textit{number of squares} relation \cite{Schmidt2019}, \textit{i.e.} $G_{xx} = G\frac{\log(r_2/r_1)}{2\pi}$, and then was normalized by the quantum of conductance $G_0$ for consistency. The final results are shown in Figs. \ref{fig:3}\textbf{A,B}, where the thermal conductivity $K/\kappa_0 T_e$ (right axis) together with the electrical conductivity (left axis) are displayed versus the partial filling factor $|\nu^{*}|$ at an electron temperature of 20 mK. Conversely, Figs. \ref{fig:3}\textbf{C,D} show the reproduced data from Fig. 4b,c of Ref. \cite{Melcer2024} taken at a reported temperature of 10 mK. 

We first note that both works point towards \textit{a priori} the same conclusion that the bulk of FQH states remains thermally conductive while the states are electrically insulating. Secondly, in terms of the Wiedemann-Franz law, which states that the ratio $\kappa_{xx}/G_{xx} = \kappa_{0}/G_0 = L_0$ is provided by a constant corresponding to the Lorenz number $L_0 = \frac{\pi^2 k_B^2}{3e^2}$, both works are in agreement and show a large violation with  $\kappa_{xx}/G_{xx}\sim\!60\ L_0$ in Ref. \cite{Melcer2024} and $(K/T_e)/G_{xx}\sim\!2.8 \times 10^5\ L_0$ in our work. We stress, however, that we opted to present the right axis of  Figs. \ref{fig:3}\textbf{A,B} in terms of the unitless ratio $K/(\kappa_0 T_e$) that does not depend on the system size unless, as we will discuss further below, heat transport becomes dominated by phonons.
	
Phonon-dominated heat transport offers a pathway to understand the substantial quantitative difference in Wiedemann-Franz law violation observed between both experiments. Based on the $T^4$ temperature dependence of $K$ above 50 mK in this and previous works \cite{Schmidt2017,Schmidt2019}, it was established that the thermal transport mechanism is indeed phonon-emission dominated. When this is the case, the thermal conductance $K$ should then scale with the 2DEG area \cite{price1982,Appleyard1998,Schmidt2019,Mittal1996,karpus1988}, \textit{i.e.} the thermal conductivity $k_{\mathrm{ph}}$ should now be given by $k_{\mathrm{ph}} = K/A_{\mathrm{2DEG}}$, where $A_{\mathrm{2DEG}}$ is the area of the 2DEG and $k_{\mathrm{ph}}$ here refers explicitly to a  thermal conductivity dominated by phonon emission. Within that scenario, and to compare our results on a equal footing with those of Ref. \cite{Melcer2024} even though at base temperature ($T=10~\mathrm{mK}$) it likely was not in the phonon-emission dominated regime, we obtain that the thermal conductivities should differ by a factor given by the ratio of the active areas in both devices, {\it i.e.} $2.95\ \mathrm{mm}^2/0.000 830\ \mathrm{mm}^2\sim\! 3500$. This is roughly the difference in magnitude between $\kappa_{xx}/\kappa_0$ and $K/{\kappa_{0}T_e}$ observed in Fig. \ref{fig:3}, and for example at filling factor  $\nu^*=0$ in the 5/2 FQH state, we obtain a value of $\left(K/{\kappa_{0}T_e}\right)/\left(\kappa_{xx}/\kappa_0 \right)\approx 8500$, a ratio that is on the same order as the ratio of probed area in the two experiments, \textit{i.e.} $\sim\!3500$. 

While both experiments capture the thermal conductance properties of the FQH bulk of the 5/2 and 7/3 filling factor, there are also notable differences between the two data sets. As seen in Fig. \ref{fig:3}\textbf{A} and within error, the thermal conductivity in the Corbino device follows qualitatively the electronic conductance and is symmetric with respect to the center of the FQH state at $\nu=5/2$. This contrasts with the trend observed in Melcer \textit{et al.}'s \cite{Melcer2024}, although a broader range of $\nu^*$ was investigated in their work. We note that over the range of $\nu^*$ that was thermally probed in our experiment, the electronic conductance is symmetric with respect to $\pm \nu^*$ (at 5/2) as one would expect for a particle-hole symmetric FQH state. This differs from the work of Melcer \textit{et al.}, where a clear asymmetry is observed in both electrical and thermal conductivity around 5/2. Future work is certainly needed to provide a clear explanation in terms of the many-body wavefunction and the possible underlying topological order.

Another noteworthy difference is the near independent behavior of $\kappa_{xx}/\kappa_{0}$ with respect to $\nu^*$ in Fig. \ref{fig:3}\textbf{C}, whereas in our case, a strong dependence is observed. Altogether, however, we stress that both experiments unambiguously demonstrate the thermally conductive nature of the bulk at the 5/2 and 7/3 FQH states leading to a large violation of the Wiedemann-Franz law.
 
\par \textit{Conclusion.} 
We presented bulk thermal conductance measurements of the 5/2 and 7/3 FQH states down to an electron temperature of 20 mK which to our knowledge is a first in the Corbino geometry. We compare our findings with a recent study by Melcer and co-workers \cite{Melcer2024} and conclude consistently that the bulk thermal conductivity is finite in the 5/2 and 7/3 FQH states at low temperature even though the bulk is electrically insulating. Accounting for the large area of our Corbino device, and in agreement with Ref. \cite{Melcer2024}, we observed a large violation of the Wiedemann-Franz law by approximately two to three orders of magnitude. Looking forward, we hope that these bulk thermal measurements will deepen our understanding of the 5/2 and 7/3 FQH states and pave the way for further thermodynamic studies in other many-body two-dimensional electron systems that could host non-trivial topological properties.

\textit{Acknowledgments.} This work has been supported by NSERC (Canada), FRQNT-funded strategic clusters INTRIQ (Québec) and Montreal-based CXC. The Princeton University portion of this research is funded in part by the Gordon and Betty Moore Foundation’s EPiQS Initiative, Grant GBMF9615.01 to Loren Pfeiffer. F. Boivin acknowledges the financial support from the FRQNT Graduate Scholarship Fund. Sample fabrication was carried out at the McGill Nanotools Microfabrication facility. We thank B.A. Schmidt and K. Bennaceur for their technical expertise during the fabrication of the Corbino sample and the earlier characterization and R. Talbot, R. Gagnon, and J. Smeros for technical assistance.\\

\textit{Corresponding author information.} Guillaume Gervais, gervais@physics.mcgill.ca.\\

\pagebreak

\widetext
\newpage
\begin{center}
	\textbf{\large Supplementary Material}
\end{center}

\setcounter{equation}{0}
\setcounter{figure}{0}
\setcounter{table}{0}
\setcounter{page}{1}
\makeatletter
\renewcommand{\theequation}{S\arabic{equation}}
\renewcommand{\thefigure}{S\arabic{figure}}
\renewcommand{\bibnumfmt}[1]{[S#1]}
\renewcommand{\citenumfont}[1]{S#1}

\doublespacing

	\section {Conductance Measurement Circuits}
	\label{Conductance Measurement}
	
\begin{figure}[!ht]
		\centering
		\includegraphics[width=\textwidth]{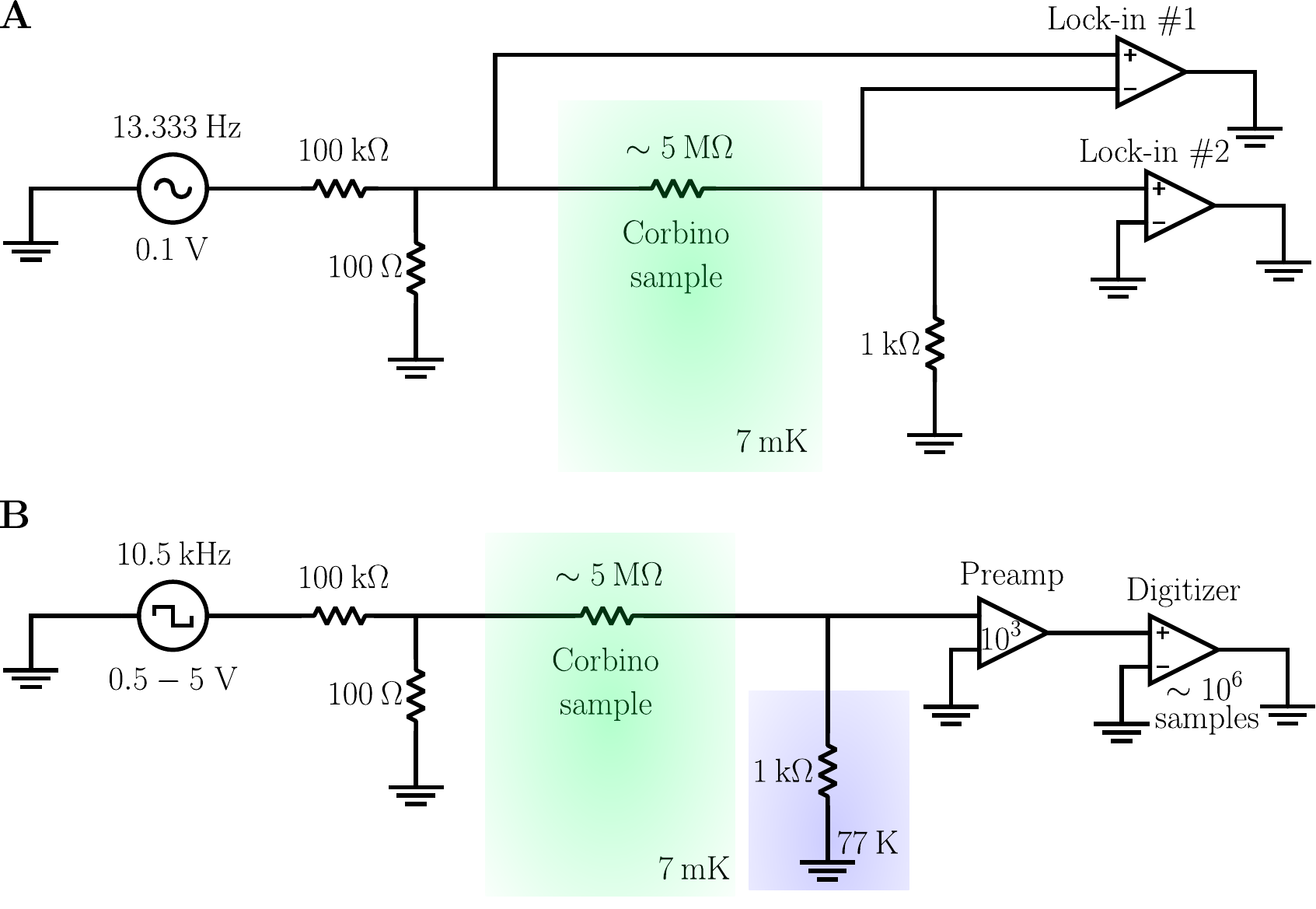}
		\caption{(\textbf{A}) Circuit used for two-point conductance measurement (transport). (\textbf{B}) Circuit used for thermal conductance measurements.}
		\label{fig:measurement_circuits}
	\end{figure}

	In Fig. 1A of the manuscript, the inset provides a simplified description of the measurement circuit for the magneto-conductance with an emphasis on the physical dimensions of the Corbino device. In order to provide more details, Fig. \ref{fig:measurement_circuits}A displays all the circuit components and the approximate resistance between the rings of the Corbino device at a magnetic field corresponding to the 5/2 fractional quantum Hall (FQH) state at the lowest temperature. Lock-in \#1 and 2 are two SR830 lock-in amplifier units synced at the same frequency (13.333 Hz). In a similar fashion, Fig. \ref{fig:measurement_circuits}B details the components of the circuit used to infer thermal conductance shown in Fig. 1B of the manuscript. For this setup, a unipolar square wave excitation is generated by a waveform generator (Highland Technology T346) and the digitizer consists of a digital lock-in amplifier (Zurich HF2LI) in scope mode. The digitizer is preceded by a pre-amplifier stage (SR560) with a gain of 1000.

	The use of a 10.5 kHz unipolar square wave excitation is required to extract the thermal time constant $\tau$ as detailed in our previous work \cite{S_Petrescu2023} where the composite fermion effective mass was determined \textit{via} the specific heat. Alternatively, this scheme also allows one to extract the thermal conductance. To further improve the signal-to-noise ratio, the 1 k$\Omega$ sense resistor was immersed in liquid nitrogen, reducing the Johnson–Nyquist noise. The left part of the circuit here acts as a voltage divider (100 $\mathrm{k\Omega}$ \textbar \textbar ~100 $\Omega$) which is necessary to bring the 0.5 - 5 V range of the lock-in amplifier to a suitable range (0.5 - 5 mV) of excitation and induce Joule heating in the few tens of pW range. The two lowest excitation voltages, namely 0.5 V and 0.75 V, required 100 batches of averaging because of their lower signal amplitudes. Conversely, due to the higher signal amplitude, for the remaining 17 biases (\textit{i.e.} 1.0~$\mathrm{V}$, 1.25~$\mathrm{V}$, 1.5~$\mathrm{V}$, ..., 4.75~$\mathrm{V}$, 5.0~$\mathrm{V}$), less averaging was needed, and 50 batches were collected for averaging.

	\section {\textit{In Situ} Thermometer Calibration of the 2DEG}
	\label{In Situ Thermometer Calibration}
	
	To obtain the temperature $T_e$ of the two-dimensional electron gas (2DEG) when subjected to a heating power $P$, we make use of the temperature dependence of the conductance $G(T_e)$ at the 5/2 FQH state. As can be inferred from Fig. 1A of the manuscript, this temperature dependence is sharp enough to allow us to extract an electron temperature $T_e$ from the conductance $G(T_e)$ making use of a univariate spline interpolation. This procedure was applied throughout all the non-exact filling factors $\nu^* \neq 0$. However, it is important to note that this procedure is limited to $\sim$ 20 mK because the temperature dependence gets progressively weaker as the temperature falls below that range, and thus the thermal conductance could only be extracted for temperatures above 20 mK.  
	
	\section {Shift and Subtract Method}
	\label{Shift and Subtract Method} 
	
\begin{figure}[!ht]
		\centering
		\includegraphics[width=\textwidth]{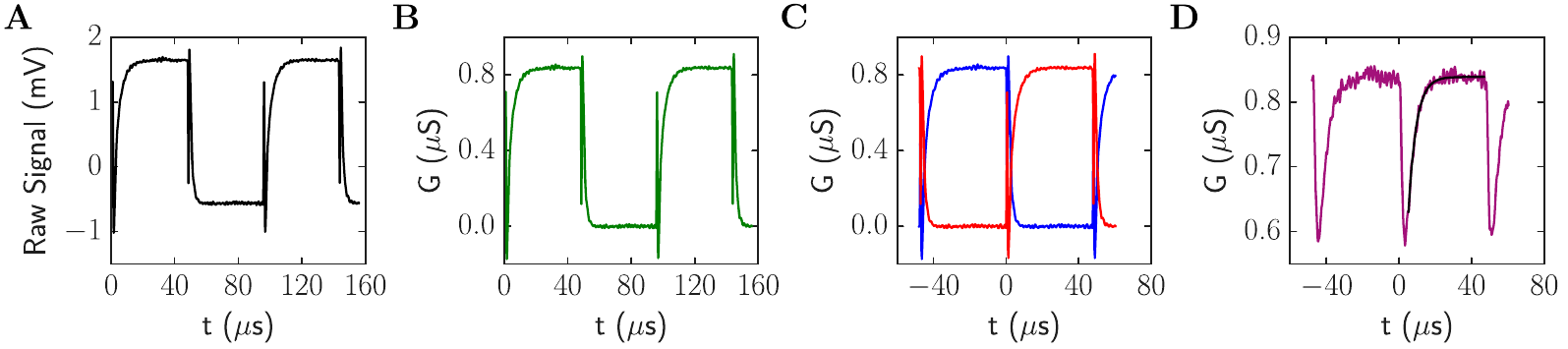}
		\caption{Example of time-resolved measurements at $\nu = 5/2$ during a square wave excitation with an amplitude of 2.5~$\mathrm{mV}$ at base temperature (7~$\mathrm{mK}$).  (\textbf{A}) Raw response of the 2DEG conductance measured by the digitizer. (\textbf{B}) Conductance of the Corbino sample calculated with the measured current as well as the voltage drop across the sample. (\textbf{C}) The conductance $G$ is shown in blue, along with the conductance offset by a half-period in red. Upon combination, the wiring resonance is subtracted out and the resulting conductance is shown in (\textbf{D}). An exponential decay fit (shown by a black line) was used to determine the final conductance.  In this example, the conductance was measured to be $G = 0.83\pm 0.01$ $\mathrm{\mu S}$.}
		\label{fig:shift_and_substract}
	\end{figure}

	When introducing a 10.5 kHz square wave, there is unavoidably spurious transients in the signal caused by the LCR resonance in the wiring of our system. To remove these unwanted resonances, a copy of the signal was numerically shifted by half a period and added to the original signal, as detailed in Fig. \ref{fig:shift_and_substract}. The conductance at equilibrium was obtained by fitting an exponential curve and was then subsequently converted to temperature $T_e$ using the temperature dependence $G(T_e)$.\\
	
	\section {Thermal Conductance}
	\label{Thermal Conductance}
	
\begin{figure}[!ht]
		\centering
		\includegraphics[width=\textwidth]{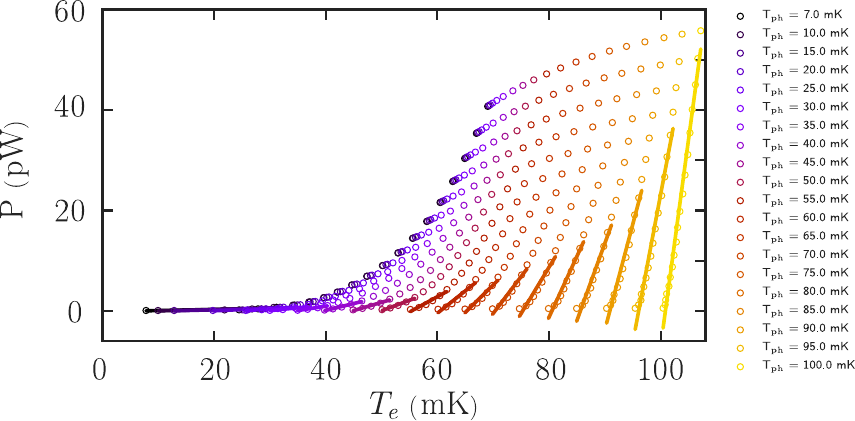}
		\caption{Power versus temperature for the 5/2 FQHS at $\nu^* = 0$. The slope of each linear fit is proportional to $K$.}
		\label{fig:K_fitting_sample}
	\end{figure}

	The procedure to extract thermal conductance was already outlined in the manuscript and follows from previous works \cite{S_Schmidt2017,S_Schmidt2019,S_Petrescu2023} in which it is explained in details. However, some minor details require specific attention here, especially regarding the range of temperatures and power used in this current work. In Fig. \ref{fig:K_fitting_sample}, all the phonon bath temperatures are shown explicitly in the legend, unlike in Fig. 1C of the manuscript where the same temperatures are shown with only a color bar. We note that a total of 19 input biases (\textit{i.e.} equivalently, 19 heating power ($P = GV_{in}^2$) data points) were collected to ensure that the temperature scale was sufficiently complete to allow for a proper interpretation of the thermal conductance.

	Additionally, we made sure that the fit from $P(T_e)$ was performed in a linear region of the $P$ versus $T_e$ data, and at low power. This is especially true for the lowest temperatures as the data can exhibit large non-linearities as $\Delta T_e$ becomes larger. Nevertheless, we required that a minimum of 4 data points were included in the fit, and Fig. \ref{fig:K_fitting_sample} shows examples of the linear fitting process.

	\section{Thermal conductance temperature dependence}
	
	In Fig. 2C,D of the manuscript, we arbitrarily shifted each curves on the log scale in order to distinguish between each filling factor investigated in our work. A least square power law fit of the form $K = AT^n$ was then performed in two regimes, \textit{i.e.} below and above 50 mK at all filling factors. This choice of constraint was based on the observation of a slight crossover in the vicinity of 50 mK indicating a different behavior in the temperature dependence of $K$. In Fig. 2C,D of the manuscript this exponent $n$ was reported for each data sets, above and below 50 mK.
	
	\section{Normalization of the thermal conductance $K$: $K/{\kappa_{0}T}$}
	The bulk thermal conductance $K$ is extracted from the relationship between the power $P$ radiated by Joule heating and the electron temperature $T_e$ of the 2DEG. Conversely, Melcer \textit{et al.}'s work \cite{S_Melcer2024} assumed a linear temperature dependence for the thermal conductivity $k_{xx} = \kappa_{xx} T$ and consequently, a thermal conductivity coefficient $\kappa_{xx}$ was reported. The following equation relates the thermal conductivity measured in Melcer \textit{et al.}'s work with the applied power
	\begin{equation}
	P = N_{\square}\frac{\kappa_{xx}(T)}{2} (T_{S}^2-T_0^2),
	\end{equation}
	where $T = \frac{T_S+T_0}{2}$ corresponds to the average electron temperature, $T_0$ is the phonon bath temperature at one end of the sample and $T_S$ is the temperature at the heated end. In, Melcer \textit{et al.}'s work, the number of squares $N_{\square}$ was estimated to be on the order of unity. In order to convert from thermal conductance $K$ data to Melcer \textit{et al.}'s factor of thermal conductivity $\kappa_{xx}$, the following equation was applied to account for the implied $T$ in Melcer \textit{et al.}'s definition,
	\begin{equation}
	\kappa_{xx} = \frac{K}{T_e} \frac{\log{(r_2/r_1)}}{2\pi}, 
	\end{equation} 
	where $\kappa_{xx}$ is the coefficient of thermal conductivity measured in our Corbino and $\frac{\log{(r_2/r_1)}}{2\pi}$ is the number of squares in the Corbino geometry. Finally, Melcer and coworkers presented their data with respect to the quantum of thermal $\kappa_0$ and electrical $G_0$ conductance. These quanta are respectively defined as
	\begin{equation}
	\kappa_0 = \frac{\pi^2 k_{B}^2}{3h} \approx 9.464 \times 10^{-13}\ \mathrm{W/K^2} \quad \text{ and } \quad  G_0 = \frac{e^2}{h}.
	\end{equation}
	Hence, our thermal conductivity ($K/\kappa_0 T$) presented in Fig. 3 of the main manuscript follows this normalization factor 
	\[ 
	\frac{K}{T_e \kappa_0} \frac{\log{(r_2/r_1)}}{2\pi},
	\]
	in order to present a fair quantitative comparison between Melcer \textit{et al.}'s work and our own work.

	\section{Estimation of the 2DEG areas}
\begin{figure}[!ht]
		\centering
		\includegraphics[width=0.5\textwidth]{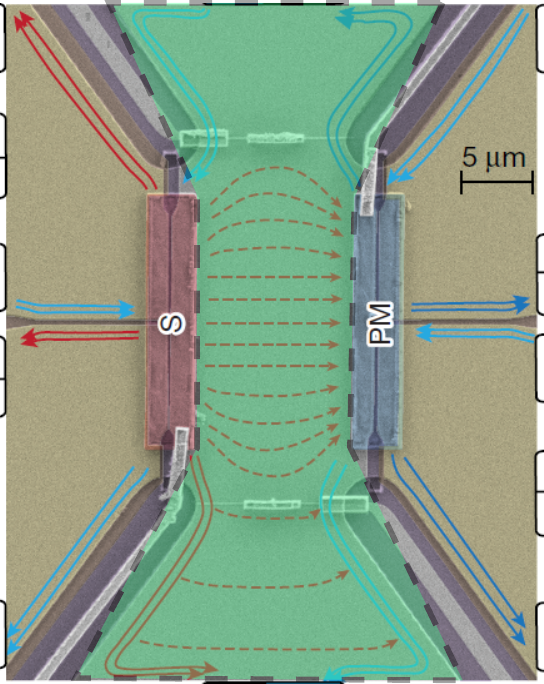}
		\caption{Estimation of the 2DEG area for the device used in \cite{S_Melcer2024}. The green shaded region corresponds to our estimation of the area of the 2DEG where the transport occurs.}
		\label{fig:MelcerDeivceArea}
	\end{figure}

	In order to determine the area of the 2DEG in the device used in Melcer \textit{et al.}'s work, an image processing software was employed to convert the number of pixels in the green shaded region (\textit{i.e.} identified as the 2DEG) shown in Fig. \ref{fig:MelcerDeivceArea} into area in units of $\mathrm{\mu m}^2$. 
	The green shaded region contains 100649 $\mathrm{pix^2}$ and the 5 $\mathrm{\mu m}$ scale bar has a length of 55 pix, yielding the following area in units of $\mathrm{\mu m}^2$
	\[ A =  (100649\ \mathrm{pix^2})\left(\frac{ 5\ \mathrm{\mu m}}{55\ \mathrm{pix}}\right)^2 \sim 831\ \mathrm{\mu m}^2.
	\]
	
	For our Corbino device, the 2DEG is contained in the annular region between the two contacts and its area is given by the surface enclosed in the ring contacts
	\[
	A = \pi (r_2^2 - r_1^2) = 2.95\ \mathrm{mm^2}.
	\]
	As a result, a ratio of $2.95\ \mathrm{mm}^2/0.000 830\ \mathrm{mm}^2\sim\! 3500$ was determined for the area probed in Mecler \textit{et al.}'s and our own work.


\begin{thebibliography}{0}%


	\bibitem{Tsui1982} D. C. Tsui, H. L. Störmer, and A. C. Gossard, \href{https://journals.aps.org/prl/abstract/10.1103/PhysRevLett.48.1559}{Phys. Rev. Lett. \textbf{48}, 1559 (1982).}
	
	
	\bibitem{Jain1989}  J.~K. Jain, \href{https://link.aps.org/doi/10.1103/PhysRevLett.63.199}{Phys. Rev. Lett. \textbf{63}, 199 (1989)}.
	
	
	\bibitem{Halperin1993} B.~I. Halperin,  P.~A. Lee, and  N. Read,  \href{https://link.aps.org/doi/10.1103/PhysRevB.47.7312}{Phys. Rev. B \textbf{47}, 7312 (1993)}.
	
	
	\bibitem{Moore1991} G. Moore, and  N. Read,  \href{https://www.sciencedirect.com/science/article/pii/055032139190407O}{Nucl. Phys. B \textbf{360}, 362 (1991)}.
	
	\bibitem{Willett1987} R. Willett, J.~P. Eisenstein, H.~L. St\"ormer, D.~C. Tsui, A.~C. Gossard, and J.~H. English,   \href{https://link.aps.org/doi/10.1103/PhysRevLett.59.1776}{Phys. Rev. Lett. \textbf{59}, 1776 (1987)}. 
	
	
	\bibitem{Eisenstein1988} J. P. Eisenstein, R. Willett, H. L. St\"ormer, and D.C. Tsui, \href{https://journals.aps.org/prl/abstract/10.1103/PhysRevLett.61.997}{Phys. Rev. Lett. \textbf{61}, 997 (1988).}
	
	\bibitem{Dolev2008} M. Dolev, M. Heiblum, V. Umansky, A. Stern, and D. Mahalu, \href{https://www.nature.com/articles/nature06855}{Nature \textbf{452}, 829 (2008).}
	
	\bibitem{Samkharadze2011} N. Samkharadze, J. D. Watson, G. Gardner, M. J. Manfra, L. N. Pfeiffer, K. W. West, and G. A. Csáthy, \href{https://journals.aps.org/prb/abstract/10.1103/PhysRevB.84.121305}{Phys. Rev. B \textbf{84}, 121305(R) (2011).}
	
	\bibitem{Tiemann2012} L. Tiemann, G. Gamez, N. Kumada, and K. Muraki, \href{https://www.science.org/doi/full/10.1126/science.1216697}{Science \textbf{335}, 828 (2012).}
	
	\bibitem{Nayak2008} C. Nayak, S. H. Simon, A. Stern, M. Freedman, and S. Das Sarma,
	\href{https://journals.aps.org/rmp/abstract/10.1103/RevModPhys.80.1083}{Rev. Mod. Phys. \textbf{80}, 1083 (2008).}
	
		
	\bibitem{Cooper2009} N. R. Cooper and A. Stern, \href{https://journals.aps.org/prl/abstract/10.1103/PhysRevLett.102.176807}{Phys. Rev. Lett. \textbf{102}, 176807 (2009).}
	
	\bibitem{Yang2009} K. Yang and B. I. Halperin, 
	\href{https://journals.aps.org/prb/abstract/10.1103/PhysRevB.79.115317}{Phys. Rev. B \textbf{79}, 115317 (2009).}
	
	\bibitem{Chickering2013} W. E. Chickering, J. P. Eisenstein, L. N. Pfeiffer, and K. W. West,
	\href{https://journals.aps.org/prb/abstract/10.1103/PhysRevB.87.075302}{Phys. Rev. B \textbf{87}, 075302 (2013).}
	
	\bibitem{Liu2018} X. Liu, T. Li, P. Zhang, L. N. Pfeiffer, K. W. West, C. Zhang, and R.-R. Du,
	\href{https://journals.aps.org/prb/abstract/10.1103/PhysRevB.97.245425}{Phys. Rev. B \textbf{97}, 245425 (2018).}
	
	\bibitem{Kane1997} C. L. Kane and M. P. A. Fisher,
	\href{https://journals.aps.org/prb/abstract/10.1103/PhysRevB.55.15832}{Phys. Rev. B \textbf{55}, 15832 (1997).}
	 
	\bibitem{Banerjee2017}M. Banerjee, M. Heiblum, A. Rosenblatt, Y. Oreg, D. E. Feldman, A. Stern, and V. Umansky,
	\href{https://www.nature.com/articles/nature22052}{Nature \textbf{545}, 75 (2017).}
	
	\bibitem{Srivastav2022} S. K. Srivastav, R. Kumar, C. Spånslätt, K. Watanabe, T. Taniguchi, A. D. Mirlin, Y. Gefen, and A. Das, 
	\href{https://www.nature.com/articles/s41467-022-32956-z}{ Nat. Commun. \textbf{13}, 5185 (2022).}
		
	\bibitem{Banerjee2018} M. Banerjee, M. Heiblum, V. Umansky, D. E. Feldman, Y. Oreg, and A. Stern, 
	\href{https://www.nature.com/articles/s41586-018-0184-1}{Nature \textbf{559}, 205 (2018).}
	
	\bibitem{Dutta2021}  B. Dutta, W. Yang, R. A. Melcer, H.~K. Kundu, M. Heiblum, V. Umansky, Y. Oreg, A. Stern, and D. Mross,   \href{https://www.science.org/doi/abs/10.1126/science.abg6116}{Science \textbf{375}, 193 (2021)}.
	
	\bibitem{Dutta2022} B. Dutta, V. Umansky, M. Banerjee, and M. Heiblum,  \href{https://www.science.org/doi/10.1126/science.abm6571}{Science \textbf{377}, 1198 (2022)}.
	
	\bibitem{Venkatachalam2012} V. Venkatachalam, S. Hart, L. N. Pfeiffer, K. W. West, and A. Yacoby, 
	\href{https://www.nature.com/articles/nphys2384}{Nat. Phys. \textbf{8}, 676 (2012).}
	
	\bibitem{Melcer2024} R. A. Melcer, A. Gil, A. K. Paul, P. Tiwari, V. Umansky, M. Heiblum, Y. Oreg, A. Stern, and E. Berg,
	\href{https://www.nature.com/articles/s41586-023-06858-z}{Nature \textbf{625}, 489 (2024).}
	
	\bibitem{Schmidt2015} B. A. Schmidt, K. Bennaceur, S. Bilodeau, G. Gervais, L. N. Pfeiffer, and K. W. West,
	\href{https://www.sciencedirect.com/science/article/pii/S0038109815001660}{Solid State Commun. \textbf{217}, 1 (2015).}

	\bibitem{Schmidt2017} B. A. Schmidt, K. Bennaceur, S. Gaucher, G. Gervais, L. N. Pfeiffer, and K. W. West, \href{https://journals.aps.org/prb/abstract/10.1103/PhysRevB.95.201306}{Phys. Rev. B \textbf{95}, 201306(R) (2017).}

	\bibitem{Petrescu2023} M. Petrescu, Z. Berkson-Korenberg, S. Vijayakrishnan, K. W. West, L. N. Pfeiffer, and G. Gervais,  \href{https://www.nature.com/articles/s41467-023-42986-w}{Nat. Commun. \textbf{14}, 7250 (2023).} 

	\bibitem{Schmidt2019}  B.~A. Schmidt, \textit{Specific heat in the fractional quantum Hall regime}, Ph.D. Thesis, \href{https://escholarship.mcgill.ca/concern/parent/ff365937v/file_sets/z029p9053}{McGill University (2019).}

	\bibitem{price1982} P. J. Price, \href{https://pubs.aip.org/aip/jap/article/53/10/6863/11514/Hot-electrons-in-a-GaAs-heterolayer-at-low}{J. Appl. Phys. \textbf{53}, 6863 (1982).}
	
	\bibitem{karpus1988} V. Karpus, Sov. Phys. Semicond. \textbf{22}, 268 (1988). [Fizh. Tekh. Poluprovodn. 22, 439 (1988)].
	
	\bibitem{Appleyard1998} N. J. Appleyard, J. T. Nicholls, M. Y. Simmons, W. R. Tribe, and M. Pepper, \href{https://journals.aps.org/prl/abstract/10.1103/PhysRevLett.81.3491}{Phys. Rev. Lett. \textbf{81}, 3491 (1998).}
	
	\bibitem{Mittal1996} A. Mittal. \textit{Electron-Phonon Scattering Rate in GaAs/AlGaAs 2DEGs Below
	0.5K.} Ph.D. Thesis, \href{https://proberlab.yale.edu/sites/default/files/files/MittalDissertation.pdf}{Yale University (1996).}
		
\end{thebibliography}

\begin{thebibliography}{0}
		
		\bibitem{S_Petrescu2023} M. Petrescu, Z. Berkson-Korenberg, S. Vijayakrishnan, K. W. West, L. N. Pfeiffer, and G. Gervais,  \href{https://www.nature.com/articles/s41467-023-42986-w}{Nat. Commun. \textbf{14}, 7250 (2023).} 
		
		\bibitem{S_Schmidt2017} B. A. Schmidt, K. Bennaceur, S. Gaucher, G. Gervais, L. N. Pfeiffer, and K. W. West, \href{https://journals.aps.org/prb/abstract/10.1103/PhysRevB.95.201306}{Phys. Rev. B \textbf{95}, 201306(R) (2017).}
		
		\bibitem{S_Schmidt2019}  B.~A. Schmidt, \textit{Specific heat in the fractional quantum Hall regime}, Ph.D. Thesis, \href{https://escholarship.mcgill.ca/concern/parent/ff365937v/file_sets/z029p9053}{McGill University (2019).}
		
		\bibitem{S_Melcer2024} R. A. Melcer, A. Gil, A. K. Paul, P. Tiwari, V. Umansky, M. Heiblum, Y. Oreg, A. Stern, and E. Berg,
		\href{https://www.nature.com/articles/s41586-023-06858-z}{Nature \textbf{625}, 489 (2024).}
		
		
	\end{thebibliography}
\end{document}